\documentclass[11pt,a4paper,twoside,openright]{article}
\usepackage{amsmath}
\usepackage{rotating}
\usepackage{graphicx}
\usepackage[T1]{fontenc} 
\usepackage[english]{babel}
\usepackage{pdflscape}
\usepackage{float}

\numberwithin{equation}{section}
\begin{document}
\vspace{1cm}

\noindent
{\bf
{\large 3D Dynamics of 4D Topological BF Theory With Boundary
}}

\vspace{.5cm}
\hrule

\vspace{1cm}

\noindent

\begin{center}{\bf  A. Amoretti, A. Blasi, N. Maggiore and N. Magnoli}
\footnote{amoretti@ge.infn.it, blasi@ge.infn.it, maggiore@ge.infn.it,
magnoli@ge.infn.it}\end{center}
\begin{center}{\footnotesize {\it
 Dipartimento di Fisica, Universit\`a di Genova \\
via Dodecaneso 33, I-16146 Genova -- Italy \\and \\INFN, Sezione di
Genova 
} }
\end{center}

\noindent
{\tt Abstract~:}
We consider the four dimensional abelian topological BF theory with a
planar boundary, following the Symanzik's method. 
We find the most general boundary conditions compatible with the
fields equations broken by the boundary. The residual gauge
invariance is described by means of two Ward identities which
generate a current algebra. We interpret this algebra
as canonical commutation relations of fields, which we use to
construct a three dimensional Lagrangian. As a remarkable by-product,
we find a (unique) boundary condition which can be read as a
duality relation between 3D dynamical variables.

\vfill\noindent
{\footnotesize {\tt Keywords:}
Boundary Quantum Field Theory,
Chern-Simons Theories,
Topological Quantum Field Theories, 
Gauge Symmetry.
\\
{\tt PACS Nos:} 
04.60.Ds	Canonical quantization
11.15.-q	Gauge field theories.
}
\newpage
\begin{small}
\end{small}

\setcounter{footnote}{0}


\section{Introduction}

In this paper we consider  the 4D BF theory, aiming at the study of 
the dynamics induced on a planar 3D boundary. The boundary is
treated 
according to the Symanzik's method \cite{syma} which 
strictly observes the Quantum Field Theory (QFT) rules of locality
and power counting, 
together with the idea of separability which is a request concerning
the propagators, $i.e.$
\begin{equation}
\Delta_{\phi_1\phi_2}(x_1,x_2) = 0 \ \ \ 
\mbox{if } x_1\cdot x_2<0,
\label{sep1}\end{equation}
where $\phi_1(x_1)$ and $\phi_2(x_2)$ are any couple of fields of the
theory. In a sense, the constraint \eqref{sep1} represents an
operative 
definition of a QFT with boundary: the boundary separates the world
into two parts in such 
a way that nothing propagates from one side to the other.

This simple idea has been very fruitful for the analysis of 
the physics on a planar boundary in several circumstances. The 
original Symanzik's motivation was the study of the Casimir effect 
\cite{syma}, but later on it has been exploited in Topological
Quantum Field Theories (TQFTs), which, as it is well known, do not
have local dynamics but their observables 
are globally defined and deal mostly with geometrical properties of
the 
manifolds they are built on \cite{Witten:1988ze,Birmingham}. A 
boundary breaks Lorentz invariance and also 
the topological character of the theory is lost; for this reason 
only in presence of a boundary a local dynamics might appear in
TQFTs. This is 
the case for the 3D Chern-Simons (CS) theory
\cite{Moore:1989yh,Blasi:1990pf,Emery:1991tf,Blasi:2008gt} which, on the
boundary, 
displays conserved chiral currents, which are the insertions of the 
fields on the two sides of the boundary obeying a Ka\v{c}-Moody (KM)
algebra whose 
central charge depends on the CS coupling constant. Similarly, the 3D 
BF theory presents, on the boundary, an algebraic structure carried
by 
two sets of   conserved chiral currents
\cite{Maggiore:1992ki,Blasi:2011pf}. 
The BF algebraic structure is more 
complex than the CS case, but we shall come to this point later. 

However, in recent years, it is in the field of condensed matter
physics that TQFTs with boundary has received much attention. This is
mainly due to the discovery in the early 1980s of a new state of
matter which cannot be described in terms of a symmetry breaking
mechanism but in terms of topological order. In contrast to the
Quantum Hall Effect (QHE), where the magnetic field breaks
time-reversal (T) symmetry, a new class of T invariant systems,
called topological insulators (TI), has been predicted
\cite{Kane:2005zz,Bernevig:2006zz} and experimentally observed
\cite{Konig:2008zz} in 3D, leading to the Quantum Spin Hall Effect
(QSHE). At the boundary of these systems, one has helical states,
namely electrons with opposite spin propagating in opposite
directions \cite{sassetti}.
The low energy sector of these materials is well described in terms
of the CS model and the BF theory. Moreover, it is important to study
these models in the presence of a boundary in order to analyze the
dynamics of the edge states. In fact, it is well known that the edge
dynamics of the QHE is successfully described by the abelian CS
theory with a boundary both in the integer and in the fractional
regimes \cite{Wen:1995qn}. In the case of 3D TI an abelian doubled CS
\cite{Bernevig:2006zz}, which is equivalent to a 3D BF theory with
cosmological constant \cite{Maggiore:1992ki}, has been introduced
\cite{Blasi:2011pf,cho}. Moreover, the authors of \cite{cho} argued
that the abelian 4D BF theory with a boundary could describe some
features of TI in 3D, in particular their edge fermionic degrees of
freedom.

\vspace{0.5cm}The paper is organized as follows: in Section 2 the
bulk 4D BF theory is described. The field equations and the Ward
identities describing the residual gauge invariance typical of the
axial gauge are derived. In Section 3 the boundary is introduced
following the Symanzik's approach. The most general boundary action
and the corresponding boundary conditions on the fields are written,
together with the Ward identities of the theory, modified by the
boundary. In Section 4 the boundary algebra is computed, which gives
rise to a surprising electromagnetic structure on the boundary, which
is to be understood both from the geometrical ($x_3=0$) and the field
theoretical (mass-shell) point of view. An ``electric'' scalar and a
``magnetic'' vector potential are identified, by means of which we
write a 3D boundary Lagrangian. A remarkable  ``duality'' relation
between the electromagnetic potentials is obtained. All the results
are summarized and discussed in detail in Section 5, and the
propagators of the full theory, including the boundary, are
explicitly given in the Appendix.

\section{The classical theory}
\label{classt}
In the abelian case the action of the four-dimensional BF model
\cite{Birmingham, Horowitz}, which describes the interaction between
the two-form $B_{\mu \nu}$ and the gauge field $A_{\mu}$, is given by:
\begin{equation}
\label{bf4}
 S_{bf}= \frac{\kappa}{2} \int d^4x \epsilon^{\mu \nu \rho \sigma}
F_{\mu \nu} B_{\rho \sigma},
\end{equation}
where $F_{\mu \nu}= \partial_{\mu} A_{\nu}-\partial_{\nu} A_{\mu}$.
It is well known that the BF theories, in any spacetime dimensions,
both in the abelian and non-abelian case, do not depend on any
coupling constant. Here, 
$\kappa$ is a constant which we have introduced in
order to distinguish the boundary terms from the bulk terms; it can
be eventually put equal to one at the end of the computation. 

The action \eqref{bf4} is invariant under the symmetries:
\begin{equation}
\label{sss1}
 \begin{split}
  &\delta^{(1)}A_{\mu}=-\partial_{\mu} \theta\\
  &\delta^{(1)}B_{\mu \nu}=0
 \end{split}
\end{equation}
and
\begin{equation}
\label{sss2}
 \begin{split}
  &\delta^{(2)} A_{\mu}=0\\
  &\delta^{(2)} B_{\mu \nu}= -(\partial_{\mu} \phi_{\nu} -
\partial_{\nu} \phi_{\mu}),\\
 \end{split}
\end{equation}
where $\theta$ and $\phi_{\mu}$ are local parameters.\\
We remind that the action \eqref{bf4} is not the most general one
compatible with the symmetries \eqref{sss1} and \eqref{sss2}. Indeed,
a Maxwell term $\int d^4xF_{\mu\nu}F^{\mu\nu}$ could be added,
coupled to an additional parameter. In this paper, we consider the
action \eqref{bf4} alone, because we are interested in the 3D
dynamics on the edge of a TQFT. We are allowed to do that, if we
think of the action \eqref{bf4} as the abelian limit of the
non-abelian 4D BF theory \cite{maggiore1}, which, as any other TQFT,
is protected from the occurrence of non-topological terms by an
additional symmetry, called ``vector supersymmetry'' in
\cite{Guadagnini:1990br}. Nevertheless, the non-abelian case,
which is much richer from the fields theoretical point of view
\cite{Batalin:1981jr,Batalin:1984jr}, and the addition of a Maxwell
term are interesting extensions which we shall present elsewhere
\cite{progress}.

Lorentz invariance will be broken by the introduction of a planar
boundary. Consequently, a convenient choice for the gauge conditions
on the two fields involved, is the axial one:
\begin{equation}
 \begin{split}
  &A_3=0\\
  &B_{i3}=0,
 \end{split}
\end{equation}
where latin letters run over 0,1,2. The axial choice is implemented
by adding to the action \eqref{bf4} the gauge fixing term
\begin{equation}
\label{gfgfgf}
 S_{gf}= \int d^4x \{b A_3 + d^i B_{i3} \},
\end{equation}
where $b$ and $d^i$ are respectively the Lagrange multipliers for the
fields $A_3$ and $B_{i3}$.
As usual, in the abelian case the ghost fields are decoupled from the
other fields. One of the main differences with the non-abelian case,
is the structure of the gauge fixing term, which, because of the
reducible symmetry \eqref{sss2}, involves ghosts for ghosts, and
therefore is highly non trivial \cite{Batalin:1981jr,Batalin:1984jr}.

Summarizing, the classical action is given by
\begin{equation}
\label{zc}
\begin{split}
 \Gamma_c[J_{\phi}]= \int d^4x \{& \kappa \epsilon^{ijk}[2\partial_i
A_j B_{k3}+ (\partial_i A_3-\partial_3 A_i)B_{jk}]+
bA_3+d^iB_{i3}+\\
&+J^{ij}_{B_{ij}}B_{ij}+2J^{i3}_{B_{i3}}B_{i3}+J^i_{A_i}A_i+J_{A_3}A_3+J_bb+J_{d_i}^id_i
\},
\end{split}
\end{equation}
where $J_\Phi$ are the external sources coupled to the quantum fields
$\Phi$.
From \eqref{zc}, we get the bulk field equations:
\begin{equation}
\label{motobf4}
 \begin{split}
  &J^i_{A_i}+\kappa \epsilon^{ijk} [2 \partial_j B_{k3}+\partial_3
B_{jk}]=0\\
  &J^{ij}_{B_{ij}}+\kappa
\epsilon^{ijk}(\partial_kA_3-\partial_3A_k)=0\\
  &J^3+b-\kappa \epsilon^{ijk} \partial_i B_{jk}=0\\
  &2J^{i3}_{B_{i3}}+d^i+2 \kappa \epsilon^{ijk} \partial_j A_k=0\\
  &A_3+J_b=0\\
  &B^{i3}+J^i_d=0.
 \end{split}
\end{equation}
It is well known \cite{Bassetto:1991ue}, that the axial gauge is not
a complete gauge fixing. A residual gauge invariance remains
on the plane $x_3=0$, which is described by two (one for each
gauge symmetry \eqref{sss1} and \eqref{sss2}) local Ward identities: 
\begin{equation}
\label{wwf1}
 W(x)\Gamma_c[J_{\phi}]=\partial_i J^i_{A^i}+ \partial_3
J^3_{A^3}+\partial_3 \frac{\delta \Gamma_c}{\delta J_b}=0,
\end{equation}
\begin{equation}
\label{wwf2}
W^i(x)\Gamma_c[J_{\phi}]=\partial_jJ^{ij}_{B^{ij}}+\partial_3J^{i3}_{B^{i3}}+\frac{1}{2}\partial_3
\frac{\delta \Gamma_c}{\delta J_{d^i}^i}=0.
\end{equation}
In the following Table 1, we list the canonical mass dimensions of
the fields of the theory:
\begin{table}[H]
 \centering
   \begin{tabular}{|c|c|c|c|c|}
\hline
&$A_{\mu}$&$B_{\mu\nu}$&$b$&$d^i$ \\ \hline
Dim&1&2&3&2 \\ \hline
 \end{tabular}
\caption{Canonical mass dimensions of the quantum fields.}
\end{table}

\section{The boundary}

To introduce a planar boundary in the theory, we adopt the Symanzik's
method \cite{syma}, which basically consists in writing the most
general boundary Lagrangian, according to the general QFT principles
of locality and power counting, and then computing, for the
modified theory, the propagators, on which the constraint of 
``separability'' is imposed. This corresponds to asking that the
propagators between points on opposite sides of the boundary
$x=0$, vanish:
\begin{equation}
\Delta_{\phi_1\phi_2}(x_1,x_2)=0\ \ \ \mbox{if}\ x_1\cdot x_2<0,
\label{sep}
\end{equation}
where $\phi_1(x_1)$ and $\phi_2(x_2)$ are two generic fields of the
theory. No other assumption is
required. In particular, as we shall see, no boundary condition is
imposed. Rather, we shall find out which are the most general ones
compatible with the presence of a separating boundary.

The most general boundary Lagrangian must respect locality, power
counting, and covariance in the plane $x_3=0$. An additional, more
subtle, constraint comes from the fact that the field equations
\eqref{motobf4} appear to be first order differential equations.
This originates from the BF action \eqref{bf4}, which contains only
first derivatives. Hence, in order to be solved, only one boundary
condition on any field is needed, and no more than one is allowed.
This obvious consideration translates into the request that the
boundary term in the action should not contain differentiated fields
(with respect to $x_3$).
In the present case, only a term proportional to
$\delta(x_3)A^i\partial_3A_i(x)$ would matter, resulting in a
boundary condition on the differentiated fields
$\left.\partial_3A_i(x)\right|_{x_3=0}$, in addition to the one on
the undifferentiated fields $\left.A_i(x)\right|_{x_3=0}$, which is
already present. This would be incompatible with the first order
differential equations \eqref{motobf4}. Hence, such a term will not
be considered.
Two more remarks on this
point: the first on the fact that this is peculiar of the 4D (or higher) since in the
lower dimensional cases (CS and 3D BF) such a term violates the power
counting constraint. The second concerns the possible presence in the
bulk of a Maxwell term, which, being of second order in the
derivatives, would spoil the previous argument, and the only possible
term we are talking of, namely
$\delta(x_3)A^i\partial_3A_i(x)$, should be in that case
included.

Summarizing, the most general boundary Lagrangian is: 
\begin{equation}
\label{lbx3}
 \mathcal{L}_{bd}=\delta(x_3) \Big[ 
a_1 A_i \tilde{B}^i +
a_2\frac{m}{2}A_iA^i +
a_3b +
\frac{a_4}{2} \epsilon^{ijk} \partial_i A_j A_k +
a_5d_iA^i
\Big],
\end{equation}
where $\tilde{B}^i\equiv\epsilon^{ijk}B_{jk}$ (and correspondingly
$J^{ij}_{B^{ik}}=\epsilon^{ijk}J_{\tilde{B}^k}$), and $a_\alpha,
\alpha = 1,\dots,5$ are constant parameters to be determined. In
order to have all the $a$-parameters massless, $a_2$ has been given
an explicit $m$-mass dependence, which will turn out to be important
in the following.

The separability condition \eqref{sep} allows for a considerable
simplification, by considering only one of the two sides of the
boundary. In fact, basically, the constraint \eqref{sep} means 
that the opposite sides of the boundary are completely decoupled. 
Hence, we can consider just one side of the boundary,
say the `+' side, forgetting about the opposite side, which can
be obtained from the `+' side by parity.

Having this in mind, the boundary Lagrangian $ \mathcal{L}_{bd}$
\eqref{lbx3} modifies the bulk field equations \eqref{motobf4} as
follows:
\begin{equation}
\label{motozrotte}
 \begin{split}
  &J^i_{A_i}+2\kappa \epsilon^{ijk}  \partial_j B_{k3}+ \alpha
\partial_3
\tilde{B}^i=-\delta(x^3)[a_1\tilde{B}^{i+}+a_2mA^{i+}+a_3b^+\\
   &
+a_4 \epsilon^{ijk}(\partial_j A_k)^++a_5d^{i+}]\\
  &\epsilon^{ijk}J_{\tilde{B}^k}+\kappa
\epsilon^{ijk}(\partial_kA_3-\partial_3A_k)=-a_1\delta(x^3)
\epsilon^{ijk}A_k^+\\
  &J^3+b-\kappa  \partial_i \tilde{B}^i=0\\
  &2J^{i3}_{B_{i3}}+d^i+2 \kappa \epsilon^{ijk} \partial_j A_k=0\\
  &A_3+J_b=-\delta(x^3)a_3\\
  &B^{i3}+J^i_d=-\delta(x^3)a_5A^{i+},
 \end{split}
\end{equation}
where the apex $+$ denotes the insertions of the fields of the theory
on the `+'-side of the boundary $x_3=0$.\\
Consequently, the boundary term $ \mathcal{L}_{bd}$ \eqref{lbx3}
breaks the local Ward identities: 
\begin{equation}
\label{wwr1}
\partial_i J_{A_i}^i+ \partial_3 J_{A_3}^3 + \partial_3 b =
- \delta(x^3) [
a_1  \partial_i \tilde{B}^{i+} + a_2m \partial_i A^{i+} +a_5
\partial_i d^{i+}
],
\end{equation}
\begin{equation}
\label{wwr2}
\epsilon^{ijk}\partial_j J_{\tilde{B}^k}+ \partial_3 J_{B_{i3}}^{i3}+
\frac{1}{2} \partial_3 d^i=-\delta(x^3)a_1 \epsilon^{ijk} \partial_j
A^+_k.
\end{equation}
We stress that the breaking terms at the r.h.s. of \eqref{wwr1} and
\eqref{wwr2} are linear in the quantum fields, and hence a
non-renormalization 
theorem ensures that they are present at the
classical level only, and do not acquire quantum corrections
\cite{brs}. For this reason, linearly broken symmetries are perfectly
allowed by general principles of QFT.

Postulating that $b(x_3=\pm \mathcal{1})=d^i(x_3=\pm \mathcal{1})=0$,
the Ward identities \eqref{wwr1} and \eqref{wwr2} can be integrated
as follows:
\begin{equation}
\label{wwrg1}
\int_{-\mathcal{1}}^{\mathcal{1}}dx_3 \partial_i J_{A_i}^i=-[
a_1 \partial_i \tilde{B}^{i+} + 
a_2m \partial_i A^{i+} +
a_5 \partial_i d^{i+}
],
\end{equation}
\begin{equation}
\label{wwrg2}
\int_{-\mathcal{1}}^{\mathcal{1}}dx_3 \epsilon^{ijk} \partial_j
J_{\tilde{B}^k}=-a_1 \epsilon^{ijk} \partial_j A^+_k.
\end{equation}

\subsection{The boundary conditions}

When dealing with a boundary, the question arises of the boundary
conditions on the fields and their derivatives. Symanzik's approach
allows to find out the most general ones, without imposing any
additional constraint. In
fact, the separability condition \eqref{sep} and the boundary
Lagrangian \eqref{lbx3} can be used to calculate the propagators
everywhere, in the bulk and on its boundary. Studying the limit of
the propagators on the different sides of the boundary, one can infer
which are the possible boundary conditions on the fields.

This program has been carried out for the 4D scalar QFT
\cite{syma,Blasi:1990pf}, for the CS theory
\cite{Blasi:1990pf,Emery:1991tf} and for the 3D BF model
\cite{Maggiore:1992ki}. But it might happen that the explicit
calculation of the propagators of the whole theory (bulk and
boundary) may become very difficult (this is the case for the 3D
Maxwell-Chern-Simons theory with boundary \cite{maxw}, for instance).
Nonetheless, even in that case it is possible to find out the
boundary conditions and the physics on the boundary (typically the
existence of conserved currents and the algebra they form), avoiding
the explicit calculation of the propagators.

In this paper, we are mainly interested in the dynamics of the 3D
boundary of the 4D BF theory, so we proceed directly to determine
which are the possible boundary conditions which characterize the
physics
on the boundary. The complete propagators of the full theory are
given in the Appendix.

In order to find out the most general boundary conditions, we
integrate the broken equations of motion \eqref{motozrotte} in an
infinitesimal interval around $x_3=0$ \cite{maxw}. Because of the
separability condition \eqref{sep}, we get the following algebraic
system, involving quantities lying on the `+' side of the boundary (the
opposite one being obtained by parity), $i.e.$ the fields on that side
of the boundary and the $a$-parameters: 
\begin{equation}
\label{sistemacondizio}
\begin{split}
&(\kappa +a_1) \tilde{B}^{i+}=
-a_2mA^{i+}
-a_3b^+
-a_4 \epsilon^{ijk}\partial_j A_k^+-a_5d^{i+}\\
&(\kappa-a_1)A^+_i=0\\
&a_3=0\\
&a_5A^{i+}=0.
\end{split}
\end{equation}
The solutions of the system \eqref{sistemacondizio} are acceptable if
the r.h.s. of  the Ward identities \eqref{wwrg1} and
\eqref{wwrg2} do not vanish. Indeed, in such a case, differentiating 
the Ward identity \eqref{wwrg1} (or \eqref{wwrg2})
with respect to $J_A$ (or $J_{\tilde{B}}$), we obtain the
inconsistency
\begin{equation}
\int dx_3\partial \delta^{(3)}(x-x')=0\ .
\end{equation}
The interesting fact we learn from this simple remark, is that the
boundary Lagrangian $\mathcal{L}_{bd}$ \eqref{lbx3} acts as a kind
of gauge-fixing for the residual gauge invariance, in the sense that
its presence is necessary (and sufficient) to calculate the
propagators on the boundary.

We see that, if we ask that the boundary term of the Ward identity
\eqref{wwrg2} does not vanish, we must impose the condition $A^{i+}
\ne 0$. Consequently, it must be:
\begin{equation}
 \begin{split}
  &a_5=0\\
  &a_1=\kappa.
 \end{split}
\end{equation}

Having done that, the system \eqref{sistemacondizio} reduces to the
single equation:
\begin{equation}
\label{sistemacondizio1}
 2 \kappa \tilde{B}^{i+}=-a_2mA^{i+}-a_4 \epsilon^{ijk}\partial_j
A_k^+,
\end{equation}
which has four different algebraic solutions, listed in Table 2.
\begin{table}[H]
\centering
\begin{tabular}{|c|c|c|c|c|}
\hline
&$a_2$&$a_4$&$A^{i+}$&$\tilde{B}^{i+}$ \\ \hline
1&$\neq 0$&$0$&$\neq 0$&$-\frac{a_2 m}{2 \kappa}A^{i+}$ \\ \hline
2&0&0&$\neq 0$&0 \\ \hline
3&$0$&$\neq 0$&$\neq 0$&$-\frac{a_4}{2
\kappa}\epsilon^{ijk}\partial_jA_k^+$ \\ \hline
4&$\neq 0$&$\neq 0$&$\neq 0$&$-\frac{1}{2 \kappa}(a_2 m A^{i+}+a_4
\epsilon^{ijk} \partial_j A^+_k)$ \\ \hline
\end{tabular}
\caption{Algebraic solutions of the equation \eqref{sistemacondizio1}}
\label{tabbu}
\end{table}
Now we notice that solutions 2 and 3 lead to ill-defined Ward
identities (the
r.h.s. of \eqref{wwrg1} vanishes), and, for this reason, they are not
acceptable.\\

We are then left with two sets of boundary conditions which satisfy
the system \eqref{sistemacondizio1}:

\begin{table}[H]
\centering
\begin{tabular}{|c|c|c|c|c|}
\hline
&$a_2$&$a_4$&$A^{i+}$&$\tilde{B}^{i+}$ \\ \hline
$I$&$\neq0$&$0$&$\neq0$&$-\frac{a_2 m}{2\kappa}A^{i+}$ \\ \hline
$II$&$\neq0$&$\neq0$&$\neq0$&$-\frac{1}{2 \kappa}(a_2 m A^{i+}+a_4
\epsilon^{ijk} \partial_j A^+_k)$ \\ \hline
\end{tabular}
\caption{Acceptable solutions of the equation
\eqref{sistemacondizio1}}
\label{tabbu1}
\end{table}
In both cases the parameter $a_2$ must be different from
zero: the presence of the massive term $m$ is necessary in order to
make the theory consistent. In other words, it is necessary that the
boundary Lagrangian is not scale-invariant. This fact is very
important for the study of the physics on the boundary, as we shall
see later.

\section{The algebra and the physics on the boundary}

In this section we derive the algebra of local observables which
is generated on the boundary, due to the residual gauge invariance of
the theory, functionally described by the Ward identities
\eqref{wwf1} and \eqref{wwf2}.\\
Next, we shall argue that it is possible to describe the physics on
the boundary in terms of two fields: a gauge field $\zeta ^i(X)$ and
a scalar massless field $\Lambda(X)$ (we recall that $X\equiv
(x_0,x_1,x_2)$, while $x\equiv (X,x_3)$). We shall identify the
Lagrangian which describes the physics on the boundary by
interpreting the boundary algebra as a set of canonical commutation
relations for the fields $\zeta^i$ and $\Lambda$.

\subsection{The boundary algebra}

Despite the fact that the solutions listed in Table 3 appear to be
different, and depending on free parameters, the broken  Ward
identities for
both of them are:
\begin{eqnarray}
 \int_{-\mathcal{1}}^{\mathcal{1}}dx_3 \partial_i J^i_{A^i} &=&\kappa
\partial_i \tilde{B}^{i+}  \label{aa1}\\
 \int_{-\mathcal{1}}^{\mathcal{1}}dx_3 \epsilon^{ijk} \partial_j
J_{\tilde{B}^k} &=&-\kappa \epsilon^{ijk} \partial_j A^+_k.
\label{aa2}
\end{eqnarray}
We therefore remark that, as a matter of fact, a unique solution
exists, which does not depends on any free parameter. We recall that
the constant $\kappa$ was introduced in order to keep trace of the
bulk dependence, but it is not a true coupling constant. We can
therefore freely put $\kappa =1$ in what follows. Evaluating the
previous relations at vanishing sources, $i.e.$ on the
mass shell, we find that:
\begin{eqnarray}
\partial_i \tilde{B}^{i+} &=& 0\label{sornul1}\\
\epsilon^{ijk} \partial_j A^+_k &=&0.\label{sornul2}
\end{eqnarray}
We now differentiate \eqref{aa1} with respect to $J_{A}^l(x')$, with
$x'$ lying on the `+' side of the boundary $x_3=0$, obtaining:
\begin{equation}
\label{alg1}
 \delta^i_l \partial_i \delta^{(3)}(X'-X)= \partial_i \Big(
\Delta_{A_l \tilde{B}^i}(x',x) \Big)_{x_3=x'_3=0^+}.
\end{equation}
Next, we express the propagator in \eqref{alg1} as follows:
\begin{equation}
 \Big( \Delta_{A_l \tilde{B}^i}(x',x)
\Big)_{x_3=x'_3=0^+}=\theta(t-t')\langle\tilde{B}^{i+}(X)A_l^+(X')\rangle+\theta(t'-t)\langle A_l^+(X')\tilde{B}^{i+}(X)\rangle.
\end{equation}
Substituting the previous identity in \eqref{alg1}, we find:
\begin{equation}
\begin{split}
 \delta^i_l \partial_i \delta^{(3)}(X'-X)&=
\delta(t-t')\langle[\tilde{B}^{0+}(X),A_l(X')]\rangle+\\
 &+
\big(\theta(t-t')\langle\partial_i\tilde{B}^{i+}(X)A_l^+(X')\rangle+\theta(t'-t)\langle A_l^+(X')\partial_i\tilde{B}^{i+}(X)\rangle
\big).
\end{split}
\end{equation}
Remembering \eqref{sornul1}, we obtain:
\begin{equation}
\label{llll}
  \delta(t-t')[\tilde{B}^{0+}(X),A_l(X')]=\delta^i_l \partial_i
\delta^{(3)}(X'-X).
\end{equation}
For $l=1,2$, it is possible to factorize $\delta(t-t')$, finding:
\begin{equation}
\label{comm1}
[\tilde{B}^{0+}(X),A_{\alpha}(X')]_{t=t'}=
\partial_{\alpha} \delta^{(2)}(X'-X),
\end{equation}
where $\alpha$ denotes the indices $1,2$. From now on, greek letters
will denote spatial coordinates on the plane $x_3=0$.

Next, differentiating \eqref{aa1} with respect to
$J_{\tilde{B}}^l(x')$, with $x'$ lying on the `+' side of the
boundary, we get:
\begin{equation}
\label{BB}
 \partial_i \Big( \Delta_{\tilde{B}_l \tilde{B}^i}(x',x)
\Big)_{x_3=x'_3=0^+}=0.
\end{equation}
Following the same reasoning which led to \eqref{comm1}, we obtain:
\begin{equation}
 [\tilde{B}^{0+}(X),\tilde{B}_l^+(X')]_{t=t'}=0.
\end{equation}
In particular, if $l=0$ the previous commutation relation become:
\begin{equation}
\label{comm2}
 [\tilde{B}^{0+}(X),\tilde{B}_0^+(X')]_{t=t'}=0.
\end{equation}
Let us now consider \eqref{aa2}. The differentiation of this identity
with respect to $J_{A^l}(x')$, (with $x_3'=0^+$), leads to:
\begin{equation}
 \epsilon^{ijk} \Big(\Delta_{A^l A_k}(x',x) \Big)_{x_3=x'_3=0^+}=0.
\end{equation}
Taking into account \eqref{sornul2},  the previous identity yields the
following commutation relation:
\begin{equation}
\label{comm3}
 [A^+_{\alpha}(X),A^{\beta +}(X')]_{t=t'}=0.
\end{equation}
Next, differentiating \eqref{aa2} with respect to
$J_{\tilde{B}}^l(x')$, we find:
\begin{equation}
\begin{split}
 (\partial_j \delta_k^l -\partial_k \delta_j^l)&\delta^{(3)}(X'-X)=\\
 &- \partial_j \Big( \Delta_{\tilde{B}^l A_k}(x',x)
\Big)_{x_3=x'_3=0^+}+ \partial_k \Big( \Delta_{\tilde{B}^l
A_j}(x',x) \Big)_{x_3=x'_3=0^+}
 \end{split}
\end{equation}
which does not provide new commutation relations.\\
In conclusion, the commutation relations \eqref{comm1}, \eqref{comm2}
and \eqref{comm3} form the following algebra of local boundary
observables:
\begin{equation}
\label{algebras}
 \begin{split}
  &[\tilde{B}^{0+}(X),A_{\alpha}(X')]_{t=t'}=
\partial_{\alpha} \delta^{(2)}(X'-X)\\
  &[\tilde{B}^{0+}(X),\tilde{B}_0^+(X')]_{t=t'}=0\\
  &[A^+_{\alpha}(X),A^{\beta +}(X')]_{t=t'}=0,
 \end{split}
\end{equation}
which will be discussed in detail in the last section of this paper,
together with the other results.

\subsection{The physics on the boundary}

Let us consider again the broken Ward identities \eqref{aa1} and
\eqref{aa2}. As we said, they describe the residual gauge invariance
of the 4D BF theory on the planar boundary $x_3=0$. Going on the mass
shell, $i.e.$ at vanishing external sources, we find the equations
\eqref{sornul1} and \eqref{sornul2} for the fields on the boundary,
for which the conditions listed in Table 3 hold. The equations
\eqref{sornul1} and \eqref{sornul2} are easily recognized as the
electromagnetic Maxwell equations for an electric ($\leftrightarrow
\tilde{B}^{i+}$) and magnetic ($\leftrightarrow {A}_i^{+}$) field. In
other words, the 4D BF theory, which, as any other TQFT does not
have local observables and has vanishing Hamiltonian, when
dimensionally reduced on a 3D planar boundary, acquires a rich,
physical, electromagnetic structure.

But we can push this further. The conditions \eqref{sornul1}
and \eqref{sornul2} allow us to express the fields $\tilde{B}^{i+}$
and $A^{i+}$ in terms of the potentials $\Lambda$ and $\zeta^i$:
\begin{equation}
 \begin{split}
  &\partial_i \tilde{B}^{i+}=0 \; \Rightarrow \;
\tilde{B}^{i+}=\epsilon^{ijk}\partial_j \zeta_k\\
  &\epsilon^{ijk} \partial_j A^+_k=0 \; \Rightarrow \;
A^+_k=\partial_k \Lambda,
 \end{split}
\end{equation}
where $\Lambda (X)$ and $\zeta^i (X)$ have canonical dimensions zero
and one,
respectively. The fields $A_i^+$ and $\tilde{B}^{i+}$ are left
invariant by translational and gauge transformations of the
potentials as follows:
\begin{equation}
 \begin{split}
  &\delta \Lambda=c\\
  &\delta \zeta_i =\partial_i \theta ,
  \end{split}
\label{bsymm}\end{equation}
where $c$ is a constant and $\theta (X)$ is a local parameter.

Let us now consider the boundary condition $I$ in Table 3. With a
suitable choice of the free parameter $a_2$, we can rewrite this
condition in terms of the fields $\zeta^i$ and $\Lambda$:
\begin{equation}
\label{condiaz}
 \epsilon^{ijk}\partial_j \zeta_k=m\partial^i \Lambda.
\end{equation}
The massive parameter $m$ in equation \eqref{condiaz} allows a
rescaling of the fields $\zeta^i$ and $\Lambda$ as follows:
\begin{equation}
 \begin{split}
  &\Lambda \rightarrow \frac{\Lambda}{\sqrt{m}}\\
  &\zeta^i \rightarrow \sqrt{m}\ \zeta^i.
 \end{split}
\end{equation}
So, thanks to the massive parameter $m$, the rescaled fields can be
given the standard canonical dimensions of a gauge field and of a
scalar field in three space-time dimensions
($[\zeta^i]=[\Lambda]=\frac{1}{2}$). Therefore, the equation
\eqref{condiaz} becomes:
\begin{equation}
\label{arat}
 \epsilon^{ijk}\partial_j \zeta_k= \partial^i \Lambda,
\end{equation}
which is exactly the duality relation between a scalar field and a
gauge field which is required to construct massless fermionic fields
in three dimensions via the tomographic representation \cite{aratyn}.
This could be interpreted as the sign that the actual degrees of
freedom of the 3D theory obtained on the boundary are fermionic
rather than bosonic. We shall come back to this point in the
conclusive Section 5.
We now consider the solution $II$ in Table 3:
\begin{equation}
\tilde{B}^{i+}=-(a_2 m A^{i+}+a_4 \epsilon^{ijk}
\partial_j A^+_k).
\end{equation}
It is evident that, if we evaluate the previous condition on the
mass-shell, the term proportional to $a_4$ vanishes due to the
condition \eqref{sornul2}, and the previous equation is
equivalent to the boundary condition $I$:
\begin{equation}
\tilde{B}^{i+}=-a_2mA^{i+}.
\label{bc}\end{equation}
 In other words, the duality
condition \eqref{arat} always holds, and the 3D physics we are
discussing here is therefore uniquely determined.

We are now able to find a 3D Lagrangian for the fields $\zeta^i$ and
$\Lambda$ which describes the physics on the boundary and which is
compatible with the duality condition \eqref{arat}. In what follows,
we shall interpret the algebra \eqref{algebras}
as a set of canonical commutation relations for the fields $\zeta^i$
and $\Lambda$, and we shall find the corresponding Lagrangian,
doing the contrary of what is commonly done, which is to  find the
canonical variables and their commutation relations from a given
Lagrangian.

Let us consider the equation \eqref{llll} with $l=0$:
\begin{equation}
 \delta(t-t')[\tilde{B}^{0+}(X),A_0(X')]=
\delta'(t-t') \delta^{(2)}(X-X').
\end{equation}
Writing this identity in terms of the fields $\zeta^i$ and $\Lambda$,
we obtain:
\begin{equation}
 \delta(t-t')\partial'_0[\epsilon^{\alpha \beta} \partial_{\alpha}
\zeta_{\beta}(X),\Lambda(X')]=
\delta'(t-t') \delta^{(2)}(X-X'),
\end{equation}
where we have factorized the operator $\partial'_0$ on the right hand
side since it acts only on the field $\Lambda$. It is easy to see 
that \mbox{$\delta(t-t')\partial'_0=-\delta'(t-t')$} and,
consequently,
we can factor out the $\delta'(t-t')$, finding the following
commutation relation:
\begin{equation}
\label{canonical1}
 [\Lambda(X'),\epsilon^{\alpha\beta} \partial_{\alpha}
\zeta_{\beta}(X)]_{t=t'}=\delta^{(2)}(X-X').
\end{equation}
Consider then the first commutation relation in \eqref{algebras}:
\begin{equation}
 [\tilde{B}^{0+}(X),A_{\alpha}(X')]_{t=t'}=
\partial_{\alpha} \delta^{(2)}(X'-X).
\end{equation}
If we express the previous identity in terms of the fields $\Lambda$
and $\zeta^i$, we find:
\begin{equation}
 \partial_{\alpha}[\epsilon^{\alpha \beta} \zeta_{\beta}(X),
\partial'_{\gamma}\Lambda(X')]_{t=t'}=
\delta^{\eta}_{\gamma} \partial_{\eta}
\delta^{(2)}(X'-X),
\end{equation}
which yields:
\begin{equation}
\label{canonical2}
 [\epsilon^{\alpha \beta} \zeta_{\beta}(X),
\partial'_{\gamma}\Lambda(X')]_{t=t'}=
\delta^{\alpha}_{\gamma} \delta^{(2)}(X'-X).
\end{equation}
We are now ready to construct the Lagrangian. The commutation
relations \eqref{canonical1} and \eqref{canonical2} allow us to
interpret the fields 
$\Pi_{(\Lambda)}\equiv\epsilon^{\alpha \beta} \partial_{\alpha}
\zeta_{\beta}$ 
and 
$\Pi_{(\zeta)\alpha}\equiv\partial_{\alpha} \Lambda$ as the conjugate
momenta of the fields $\Lambda$ and
$\tilde{\zeta}^\alpha\equiv\epsilon^{\alpha \beta} \zeta_{\beta}$
respectively.\\
With these assumptions, the Lagrangian of the system
$\mathcal{L}=\sum \Pi\dot{\Phi}-H$,
where $H$ is the Hamiltonian of the system, is given by: 
\begin{equation}
 \mathcal{L}=\epsilon^{\alpha \beta} \partial_{\alpha} \zeta_{\beta}
\partial_t \Lambda+\partial_{\alpha} \Lambda \epsilon^{\alpha \beta}
\partial_t \zeta_{\beta}-(\epsilon^{\alpha \beta}
\partial_{\alpha} \zeta_{\beta})^2-(\partial_{\alpha}
\Lambda)^2,
\label{blagr}\end{equation}
which is equivalent to the Lagrangian postulated in \cite{cho} for
the study of the topological insulators. 
We stress that the Lagrangian \eqref{blagr} is the most general one 
compatible with power counting and respecting the symmetries 
\eqref{bsymm}\footnote{We thank the referee for suggesting us to 
clarify the issue of the uniqueness of \eqref{blagr}}. Furthermore, 
the coefficients of the terms appearing in \eqref{blagr} are fixed by 
making the field equations of motion compatible with the duality 
relation \eqref{arat}. Moreover, if we omit thekinetic term, $\mathcal{L}$ 
is equivalent to the Lagrangianconsidered in \cite{balachandran} to study 
the edge states of the 4D BF theory.\\

\section{Summary and discussion}

The main results presented in this paper are

\vspace{0.5cm}

1) {\bf boundary as gauge fixing}

 According to the Symanzik's approach, the boundary 
conditions on the fields are not imposed, but, rather, are derived
from the form of the propagators. Now, this is not 
always feasible, and almost always quite difficult (we are talking 
about computing the propagators of the theory, including the 
boundary and satisfying the separability condition \eqref{sep1}). In
a 
previous paper \cite{maxw} we already succeeded in finding out the
boundary 
conditions (which are the starting point for the study of the physics 
on the boundary) without computing explicitly the propagators. At 
the end, one is faced with a nonlinear algebraic system whose 
unknowns are the parameters on which the boundary Lagrangian depends, 
and the fields (and its derivatives) on the boundary. Most of these 
solutions are inconsistent, or unacceptable for some reasons, and in 
\cite{maxw} these unphysical solutions were ruled out one by one.
Here, 
we found a nice, general criterion to get the same result: the 
r.h.s. of the Ward identities \eqref{wiconcl1} and \eqref{wiconcl2},
$i.e.$ the 
linear breaking due to the boundary term in the action, must always 
be different from zero. Otherwise, the propagators cannot be defined. 
The nice interpretation of this statement, is that the boundary term 
in the action plays the role of a gauge fixing of the residual gauge 
invariance on the boundary. This observation leads immediately to the
solutions $I$ 
and $II$ listed in Table 3. 

\vspace{0.5cm}
2) {\bf Ward identities in presence of a boundary}

\begin{eqnarray}
 \int_{-\infty}^{+\infty}dx_3 \partial_i J^i_{A^i} &=&
\partial_i \tilde{B}^{i+}  \label{wiconcl1}\\
 \int_{-\mathcal{1}}^{+\mathcal{1}}dx_3 \epsilon^{ijk} \partial_j
J_{\tilde{B}^k} &=&- \epsilon^{ijk} \partial_j A^+_k.
\label{wiconcl2}
\end{eqnarray}

Quite remarkably, the apparently distinct solutions $I$ and $II$ of Table
3
physically coincide, since they lead to the same Ward 
identities on the boundary. This is the first evidence of the 
striking electromagnetic structure which determines the physics on 
the boundary, as we shall discuss shortly. In addition, despite the
fact that the solutions 
depend on free parameters, when put into the Ward identities
\eqref{wiconcl1} 
and \eqref{wiconcl2}, which contain all the physical information,
these disappear. The 
separability condition isolates a unique dynamics on the boundary, 
without any dependence on free parameters. 

\vspace{0.5cm}3) {\bf electromagnetism on the boundary} 
\begin{eqnarray}
\partial_i \tilde{B}^{i+}&=0 \; \Rightarrow \;&
\tilde{B}^{i+}=\epsilon^{ijk}\partial_j \zeta_k \label{max1}\\
\epsilon^{ijk} \partial_j A^+_k &=0 \; \Rightarrow \;&
A^+_k=\partial_k \Lambda\label{max2},
\end{eqnarray}

On the boundary $x_3=0$, $and$ on the mass shell 
$J_{\phi}=\left.\frac{\delta 
\Gamma_{c}}{\delta\phi}\right|_{J=0}=0$ (we stress this double
constraint defining the boundary), the 4D topological BF theory 
displays Maxwell equations for an electric field and a magnetic 
field, to be identified with the boundary insertions
$\tilde{B}^{i+}$ and $A^{+}_{i}$, respectively. This is a direct 
consequence of the result 2). Consequently, two potentials can be 
introduced: an electric scalar potential $\Lambda(X)$ and a magnetic 
vector potential $\zeta^{i}(X)$, depending on the 3D coordinates on
the plane $x_3=0$: $X=(x_0,x_1,x_2)$.

\vspace{0.5cm}4) {\bf duality} 

\begin{equation}
\label{duality}
 \epsilon^{ijk}\partial_j \zeta_k= \partial^i \Lambda.
\end{equation}
The solutions of Table 3, $i.e.$ the possible boundary 
conditions on the fields, translates in the ``duality'' 
condition between the potentials \eqref{duality}.
This confirms the fact that the dynamics on the boundary is uniquely 
determined by the Ward identities \eqref{wiconcl1} and 
\eqref{wiconcl2}. We find here, in a well defined field theoretical 
framework, a strong motivation for a relation which is known since a 
long time \cite{aratyn}, where this duality (or ``tomographic'')
relation  was introduced to 
give a Bose description of fermions in 3D. Here, this condition appears as
the unique boundary condition on the fields
$A^{i+}=\tilde{B}^{i+}$, written in 
terms of electromagnetic potentials defined by the boundary Maxwell 
equations \eqref{max1} and \eqref{max2}. This strongly suggest that
the actual 
degrees of freedom of the dimensionally reduced 3D theory are 
fermionic, confirming recent developments concerning the edge 
states of topological insulators, which seem to be described in terms 
of fermion fields \cite{Hasan:2010xy}.

\vspace{0.5cm}5) {\bf 3D boundary algebra} 

\begin{equation}
\label{algebraconcl}
 \begin{split}
  &[\tilde{B}^{0+}(X),A_{\alpha}(X')]_{t=t'}=
\partial_{\alpha} \delta^{(2)}(X'-X)\\
  &[\tilde{B}^{0+}(X),\tilde{B}_0^+(X')]_{t=t'}=0\\
  &[A^+_{\alpha}(X),A^{\beta +}(X')]_{t=t'}=0,
 \end{split}
\end{equation}
On the boundary, the above algebra is found. It is formed by a
vectorial, conserved current, whose 
3D components are the insertions of the fields on (one side of) the
boundary 
($\tilde{B}^{0+}(X)$ and  $A^+_\alpha(X)\ \alpha=1,2$, related by the 
duality-boundary condition \eqref{bc}). We stress that the
conservation of the 
current is obtained
on the mass-shell, $i.e.$ at vanishing external sources $J_\phi$.
This is in perfect analogy with what happens in the 3D CS and the 3D
BF theory. In all cases, the conservation of the currents comes from
the Ward identities of the residual gauge invariance broken by the
most general boundary term respecting Symanzik's separability
condition \eqref{sep1}, going on the mass-shell, and exploiting the
boundary condition previously found on the quantum fields. The
physical interpretation of the current conservation is different,
since in the 3D CS and BF cases, it leads, thanks to the boundary
conditions, to the chirality of the currents. In the 4D BF case the
current conservation \eqref{max1} (again, together with the
duality-boundary condition \eqref{duality}), is tightly related to
the electromagnetic structure and the consequent determination of the
electromagnetic potentials. One more comment on the algebra
\eqref{algebraconcl} is on order. The 3D BF theory shows two types of
algebraic structures on its 2D planar boundary, whether the
cosmological constant $\lambda$, whose presence is peculiar of the 3D
case, is vanishing or not \cite{Maggiore:1992ki}. For $\lambda\neq
0$, one finds the direct sum of two KM algebras, and this is not
surprising because the 3D BF theory with cosmological constant can
be written in terms of two CS theories with opposite coupling
constants, each of which shows a boundary KM algebra. On the
contrary, the algebra found for vanishing cosmological constant is
truly BF-like. Indeed in that case the BF theory, in any dimension,
cannot be rephrased in terms of CS actions. A remarkable check 
of our results, is that we find exactly the same
algebraic structure as the one found for the 3D BF case for
$\lambda=0$, whose interesting features and relationships with other
bulk theories are discussed elsewhere \cite{progress}. In a different
framework and language the same
algebra has been found in \cite{balachandran}, written in terms of
the same dynamical variables we treated in this paper, $i.e.$ the
edge states of 4D BF theory.

\vspace{0.5cm}6) {\bf canonical commutation relations and dimensional
reduction} 

\begin{eqnarray}
\left [\Lambda(X),\Pi_{(\Lambda)}(X')\right]_{t=t'} &=&
\delta^{(2)}(X-X')
\\
\left[\tilde\zeta^\alpha(X),\Pi_{(\zeta)\beta}(X')\right]_{t=t'}  &=&
\delta^{\alpha}_{\beta} \delta^{(2)}(X-X'),
\end{eqnarray}
where $\Pi_{(\Lambda)}\equiv\epsilon^{\alpha \beta} \partial_{\alpha}
\zeta_{\beta}$ and $\Pi_{(\zeta)\alpha}\equiv\partial_{\alpha}
\Lambda$ are the conjugate
momenta of the fields $\Lambda$ and
$\tilde{\zeta}^\alpha\equiv\epsilon^{\alpha \beta} \zeta_{\beta}$
respectively. The point to stress here, is that, written in terms of
the electromagnetic potentials \eqref{wiconcl1} and \eqref{wiconcl2},
the boundary algebra \eqref{algebraconcl} can be interpreted as a set
of canonical commutation relations, for the canonically conjugate
variables. Once realized this, it is almost immediate to write down
the corresponding 3D Lagrangian, which is uniquely determined by our
procedure. Indeed this analysis can be viewed as a systematic way to
find (D$-1$)-dimensional Lagrangians out of D-dimensional bulk
theories. 
It is a surprising and welcome result, that this new way of
dimensionally reducing D-dimensional theories originates from the
algebraic structure found on the boundary, interpreted as a set of
canonical commutation relations, and which comes from the Ward
identities describing the residual gauge invariance on the boundary
and broken (by the boundary itself) in the most general (and unique) 
way compatible with the Symanzik's simple criterion of
separability. 

\vspace{0.5cm}7) {\bf 3D Lagrangian} 

\begin{eqnarray}
\mathcal{L} &=& \sum \Pi\dot{\Phi}-H \nonumber \\
&=& \epsilon^{\alpha \beta} \partial_{\alpha} \zeta_{\beta}
\partial_t \Lambda+\partial_{\alpha} \Lambda \epsilon^{\alpha \beta}
\partial_t \zeta_{\beta}-(\epsilon^{\alpha \beta}
\partial_{\alpha} \zeta_{\beta})^2-(\partial_{\alpha}
\Lambda)^2\label{3dlagrangian}
\end{eqnarray}
This is the 3D Lagrangian obtained on the mass-shell boundary of the
4D topological BF theory. It is the unique solution compatible with
the QFT request of locality, power counting and with the Symanzik's
criterion of separability \eqref{sep1}. It is left invariant by gauge
and translational transformations. It is non-covariant, and its
dynamical variables (scalar and vector potentials) are coupled in a
non-trivial way. Quite remarkably, this action, uniquely derived here
by very general QFT principles, coincides with the one studied in
\cite{balachandran} for the edge states of the 4D BF theory, where,
the same algebraic origin is stressed. In a completely different 
theoretical framework, the action \eqref{3dlagrangian} is employed to
study the surface of 4D (3+1) topological insulators \cite{cho}. The
duality relation \eqref{duality} is there exploited to extract the
desired fermionic degrees of freedom.

\vspace{0.5cm}

There are several interesting developments of the results presented
in this paper.  The most obvious is the non-abelian extension, with
particular attention to the duality relation \eqref{duality}, to the
boundary algebra \eqref{algebraconcl} and to the boundary Lagrangian
\eqref{3dlagrangian}. Moreover, it is of interest to apply our
method to dimensionally reduce 5D bulk theories, in order to find out
the resulting 4D actions.

\appendix\section{The propagators}

In this appendix we shall derive the propagators of the 4D BF model
\eqref{bf4} with boundary \eqref{lbx3}, taking into account the
boundary conditions in Table 3.\\

The separability condition \eqref{sep} is satisfied by propagators
with the following form:
\begin{equation}
 \Delta_{\phi_1
\phi_2}(x,x')=\theta(x_3)\theta(x_3')\Delta^+_{\phi_1
\phi_2}(x,x')+\theta(-x_3)\theta(-x_3')\Delta^-_{\phi_1
\phi_2}(x,x'),
\end{equation}
where $\Delta^\pm_{\phi_1 \phi_2}$ are the propagators on the
`$\pm$' side of the boundary. They are solutions of the system of
equations obtained by differentiating the bulk equations of motion
\eqref{motobf4} with respect to the sources of the fields. They must
be compatible with the boundary equations of motion
\eqref{motozrotte} and with the Ward identities \eqref{wwr1} and
\eqref{wwr2}. Since $\Delta^+_{\phi_1 \phi_2}$ and
$\Delta^-_{\phi_1 \phi_2}$ are transformed into each other by a
parity transformation, in this appendix we derive a solution for
$\Delta^+_{\phi_1 \phi_2}$, where $x_3,x_3' \ge 0$. In what
follows we shall omit the index $+$.\\
If we differentiate the equations of motion \eqref{motobf4} with
respect to the sources of the fields and we evaluate the expressions
obtained at vanishing sources, we get a system of equations for the
propagators of the theory:
\begin{equation}
\label{propbf4}
 \begin{split}
 &\Delta_{A_3 \psi}(x,x')=0 \qquad \qquad \forall \; \psi(x') \neq
b(x')\\
 &\Delta_{A_3 b}(x,x')=-\delta^{(4)}(x-x')\\
 &\Delta_{B^{i3} \psi} (x,x')=0 \qquad \qquad \forall \; \psi(x')
\neq d^i(x')\\
 &\Delta_{B^{i3} d_l} (x,x')=- \delta^i_{\;l} \delta^{(4)}(x-x')\\
 &\partial_3 \Delta_{A_l
\tilde{B}^i}(x',x)=-\delta^i_{\;l}\delta^{(4)}(x-x')\\
 &\partial_3 \Delta_{\tilde{B}^l \tilde{B}_i}(x',x)=0\\
 &\partial_3 \Delta_{b \tilde{B}^i}(x',x)=0\\
 &\partial_3 \Delta_{d_l \tilde{B}^i}(x',x)=2 \epsilon^{ij}_{\;\;l}
\partial_j \delta^{(4)}(x-x')\\
 &\partial_3 \Delta_{A_l A^i}(x',x)=0\\
 &\partial_3 \Delta_{\tilde{B}_l
A^i}(x',x)=\delta^i_{\;l}\delta^{(4)}(x-x')\\
 &\partial_3 \Delta_{b A^i}(x',x)=-\partial_i \delta^{(4)}(x'-x)\\
 &\partial_3 \Delta_{d_l A^i}(x',x)=0\\
 &\Delta_{A_l b}(x',x)= \partial_i \Delta_{A_l
\tilde{B}^i}(x',x)\\ 
 &\Delta_{\tilde{B}_l b}(x',x)=\partial_i
\Delta_{\tilde{B}_l \tilde{B}^i}(x',x)\\
 &\Delta_{b b}(x',x)= \partial_i \Delta_{b
\tilde{B}^i}(x',x)\\
 &\Delta_{d_l b}(x',x)=\partial_i \Delta_{d_l
\tilde{B}^i}(x',x)\\
 &\Delta_{A_l d^i}=-2  \epsilon^{ijk}\partial_j \Delta_{A_l
A_k}(x',x)\\
 &\Delta_{\tilde{B}_ld^i}(x',x)=-2  \epsilon^{ijk}\partial_j
\Delta_{\tilde{B}_l A_k}(x',x)\\
 &\Delta_{d_l d^i}(x',x)=-2  \epsilon^{ijk} \partial_j
\Delta_{d_l A_k}(x',x)\\
 &\Delta_{b d^i}(x',x)=-2 \epsilon^{ijk} \partial_j \Delta_{b
A_k}(x',x).
 \end{split}
\end{equation}
Notice that it follows directly from the gauge conditions, $i.e.$
from the last two equations in \eqref{motobf4}, that the Green
functions containing $A_3$ and $B_{i3}$ vanish except $\Delta_{A_3
b}(x,x')=-\delta^{(4)}(x-x') $ and $\Delta_{B^{i3} d_l} (x,x')=-
\delta^i_{\;l} \delta^{(4)}(x-x')$ and, for this reason, we do not
list these propagators in the following.\\
The most general solution of the previous system is:
\begin{equation}
\label{desolp}
\begin{split}
&\Delta_{\phi_1\phi_2}(x',x)=\\
&\begin{pmatrix} 
 \Xi_l^{\;i}(X,X') & \delta^i_l T_{c_1}(x,x') & -2
 \epsilon^{ij}_{\; \; k} \partial_j \Xi_l^{\;k}(X',X) & -
\partial_l T_{c_1}(x,x')\\ 
 \; & \; & \; & \;\\
 -\delta^i_l T_{c_2}(x',x) & \Omega_l^{\;i}(X,X') &
2\epsilon^{ij}_{\; \; l} \partial_j T_{c_2}(x',x) &  \partial_i
\Omega_l^{\;i}(X,X')\\
 \; & \; & \; & \;\\
 -2  \epsilon^{ij}_{\; \; k} \partial_j \Xi_l^{\;k}(X',X) & 2
\epsilon^{ij}_{\; \; l} \partial_j T_{c_4}(x,x') & 4 
\epsilon^{ijk}\partial_j \epsilon_{kpq}\partial^p \Xi_l^{q}(X,X') & 0
\\
 \; & \; & \; & \;\\
 \partial^i T_{c_3}(x',x) &  \partial^l \Omega_l^{\;i}(X',X) &
0 &  \partial^l \partial_i \Omega_l^{\;i}(X,X')
\end{pmatrix}
\end{split}
\end{equation}

The labels $\phi_1$ and $\phi_2$ run over the set of fields $\{ A^i,
\tilde{B}^i, d^i,b \}$. $T_{c_i}(x,x')$ is the tempered distribution
$(\theta(x_3-x_3')+c_i)\delta^3(X-X')$, $\Xi_l^{\;i}(X,X')$ and
$\Omega_l^{\;i}(X,X')$ are generic functions of the transverse
coordinates $X\equiv(x_0,x_1,x_2)$, and $c_i, i=1,...,4$ are
constant parameters.\\

Let us now consider the boundary conditions $I$ and $II$ in Table 3.
Remarkably, for both these solutions the Ward identities
\eqref{wwrg1} and \eqref{wwrg2} take the following form:
\begin{equation}
\label{wibfsol}
 \begin{split}
  &\int_{-\mathcal{1}}^{\mathcal{1}}dx_3 \partial_i J^i_{A^i}=
\partial_i \tilde{B}^{i+}  \\
  &\int_{-\mathcal{1}}^{\mathcal{1}}dx_3 \epsilon^{ijk} \partial_j
J_{\tilde{B}^k}=- \epsilon^{ijk} \partial_j A^+_k.
 \end{split}
\end{equation}
Differentiating the equations \eqref{wibfsol} with respect to the
sources $J_{A^l}(x'),$ $J_{\tilde{B}^l}(x'),J_{d^l}(x')$ and
$J_b(x')$, we obtain eight differential equations for the propagators:
\begin{eqnarray}
 \partial_i \delta^{i}_{\;l} \delta^{(3)}(X'-X)
&=&
\partial_i \Big(\Delta_{A_l \tilde{B}^i}(x',x)
\Big)_{x_3=0}\label{uno}\\
 \partial_i\Big(\Delta_{\tilde{B}_l \tilde{B}^i}(x',x) \Big)_{x_3=0}
&=&0\label{due}\\
 \partial_i \Big(\Delta_{d_l \tilde{B}^i}(x',x) \Big)_{x_3=0} 
&=&0\label{tre}\\
 \partial_i \Big(\Delta_{b \tilde{B}^i}(x',x)
\Big)_{x_3=0}&=&0\label{quattro}\\
 \epsilon^{ijk} \partial_j \Big(\Delta_{A_l A^k}(x',x)
\Big)_{x_3=0}&=&0\label{cinque}\\
 \epsilon^{ijk} \partial_j \delta_{kl} \delta^{(3)}(X'-X)
&=&
- \epsilon^{ijk} \partial_j \Big(\Delta_{\tilde{B}_l A_k}(x',x)
\Big)_{x_3=0}\label{sei}\\ 
 \epsilon^{ijk}\partial_j \Big(\Delta_{d_l A_k}(x',x)
\Big)_{x_3=0}&=&0\label{sette}\\
 \epsilon^{ijk} \partial_j \Big(\Delta_{b A_k}(x',x)
\Big)_{x_3=0}&=&0.\label{otto}
\end{eqnarray}
Substituting the propagators \eqref{desolp} in the above system of
differential equations, we get the following constraints on the
parameters $c_1$ and $c_2$:
\begin{equation}
\label{psymmetric}
 \begin{split}
  &c_1=-1\\
  &c_2=0.
 \end{split}
\end{equation}
Additional constraints come from the request that the Lagrange
multipliers $b(x)$ and $d^i(x)$ vanish at $x_3\rightarrow\infty$.
Consequently, the propagators involving these fields must satisfy:
\begin{equation}
 \begin{split}
  &\lim_{x_3 \rightarrow + \mathcal{1}} 2 \epsilon^{ij}_{\; \; l}
\partial_j T_{c_4}(x,x')=0\\
  &\lim_{x_3 \rightarrow + \mathcal{1}} \partial_i T_{c_3}(x',x)=0,
 \end{split}
\end{equation}
which yield:
\begin{equation}
 \begin{split}
  &c_3=-1\\
  &c_4=0.
 \end{split}
\end{equation}
Moreover,  the propagators $\Delta_{A_l A^i}(x',x)$ and
$\Delta_{\tilde{B}_l \tilde{B}^i}(x',x)$ must be symmetric for the
exchange $\{x,i\} \leftrightarrow \{x',l\}$. As a consequence, taking
into account \eqref{due} and \eqref{sette}, we obtain that the
functions $\Xi_l^{\;i}(X,X')$ and $\Omega_l^{\;i}(X,X')$ take the
following form:
\begin{equation}
 \begin{split}
  &\Xi_l^{\;i}(X,X')=\partial^i \partial_l \eta (X-X')\\
  &\Omega_l^{\;i}(X,X')=\epsilon^{ijk} \partial_j \epsilon_l^{\;rs}
\partial_r \phi_{ks} (X-X'),
 \end{split}
\end{equation}
where $\eta(X-X')$ and $\phi_{ks}(X-X')$ are generic functions of
the transverse coordinates $X-X'$, and have canonical mass dimensions
zero and two, respectively. In terms of these functions, the matrix
of propagators \eqref{desolp} finally reads: 
\begin{equation}
\label{desolp1}
\begin{split}
&\Delta_{AB}(x',x)=\\
&\begin{pmatrix} 
 \partial^i \partial_l \eta(X-X') &
-\delta^i_l T_{-1}(x,x') & 0 & - \partial_l
T_{-1}(x,x')\\ 
 \; & \; & \; & \;\\
 -\delta^i_l T_{0}(x',x) & \epsilon^{ijk} \partial_j
\epsilon_l^{\;rs} \partial_r \phi_{ks} (X-X') & 2\epsilon^{ij}_{\;
\; l} \partial_j T_{0}(x',x) & 0\\
 \; & \; & \; & \;\\
 0 & 2 \epsilon^{ij}_{\; \; l} \partial_j T_{0}(x,x') & 0 & 0 \\
 \; & \; & \; & \;\\
 \partial^i T_{-1}(x',x) & 0 & 0 & 0
\end{pmatrix}
\end{split}
\end{equation}


\end{document}